\documentclass[twocolumn]{aastex701}

\usepackage{CJK}
\usepackage{enumitem}
\usepackage{amsmath}
\usepackage{placeins}
\newcommand{\tstars}{$\text{***}$}
\newcommand{\dstars}{$\text{**}$}
\newcommand{\ostar}{$\text{*}$}

\shortauthors{Zhang et al.}

\begin{document}
\begin{CJK*}{UTF8}{gbsn}

\title{The Intrinsic Multi‑phase Gas--Black Hole Connection across Scales in IllustrisTNG}

\correspondingauthor{Xiaoxia Zhang; Taotao Fang; Si-Yue Yu}
\email{zhangxx@xmu.edu.cn; fangt@xmu.edu.cn; syu@xmu.edu.cn}

\author[0000-0003-4832-9422]{Xiaoxia Zhang (张小霞)}  
\affiliation{Department of Astronomy, Xiamen University, Xiamen, Fujian 361005, People's Republic of China}
\email{}
\author[0000-0002-2853-3808]{Taotao Fang (方陶陶)}
\affiliation{Department of Astronomy, Xiamen University, Xiamen, Fujian 361005, People's Republic of China}
\email{}
\author[0000-0002-3940-2950]{Shulan Yan (鄢淑澜)}
\affiliation{Department of Astronomy, Xiamen University, Xiamen, Fujian 361005, People's Republic of China}
\email{}
\author[0000-0002-3462-4175]{Si-Yue Yu (余思悦)}
\affiliation{Department of Astronomy, Xiamen University, Xiamen, Fujian 361005, People's Republic of China}
\email{}
\author[0000-0003-3564-6437]{Feng Yuan (袁峰)}
\affiliation{Center for Astronomy and Astrophysics and Department of Physics, Fudan University, Shanghai 200438, People's Republic of China}
\email{}

\begin{abstract}

The relationship between supermassive black holes and the multiphase circumgalactic medium is central to understanding the co-evolution of galaxies and their central black holes. We investigate this relationship using the IllustrisTNG100 simulation with a sample of 5089 central galaxies at $z=0$, measuring the partial correlation between central black hole mass and the mass of cold ($T < 10^4$\,K), cool ($10^4 \le T < 10^5$\,K), warm ($10^5 \le T < 10^6$\,K), and hot ($T \ge 10^6$\,K) gas within $0.03R_{200}$, $0.15R_{200}$, and $R_{200}$, after accounting for stellar and dark matter halo mass. We find that after removing these confounding factors, black hole mass shows a significant negative partial correlation ($\rho \approx -0.37$) with cold gas within $R_{200}$ and $0.15R_{200}$, whereas warm and hot gas exhibit no substantial intrinsic correlation. The residual plane reveals a threshold pattern: galaxies with over-massive black holes show systematically reduced cold gas, consistent with the cumulative impact of AGN feedback. The anti-correlation persists across environments with a weak trend in local density, and varies with galaxy type (star-forming, green valley, and quenched). These results provide a quantitative multiphase diagnostic of AGN feedback in TNG and support a picture in which feedback progressively removes cold gas, offering testable predictions for future multiwavelength surveys.

\end{abstract}



\section{Introduction} 
\label{sec:intro}

The tight empirical correlations between supermassive black hole (SMBH) mass and host galaxy properties, such as bulge mass and stellar velocity dispersion, provide strong evidence for the co-evolution of galaxies and their central black holes \citep[e.g.,][]{2013ARA&A..51..511K}. The prevailing framework attributes this coupling to energetic feedback from active galactic nuclei (AGN), which regulates galaxy growth by heating, ejecting, or preventing the cooling of circumgalactic gas \citep{1998A&A...331L...1S, 2005Natur.433..604D, 2012ARA&A..50..455F}. A direct test of this picture lies in examining the relationship between SMBHs and the multiphase circumgalactic medium (CGM), which serves as both the fuel reservoir for black hole accretion and the primary recipient of feedback energy \citep[e.g.,][]{2017ARA&A..55..389T}.

Observationally, the SMBH--CGM connection has been explored one thermodynamic phase at a time. For the hot, X-ray emitting gas ($T > 10^6$\,K), scaling relations have been established across early-type galaxies, groups, and clusters \citep[e.g.,][]{2018ApJ...852..131B, 2019MNRAS.488L.134L, 2019ApJ...875..141P, 2019ApJ...884..169G}. For the cold atomic phase, recent analyses of nearby galaxies show that the \ion{H}{1} mass fraction correlates more strongly with SMBH mass than with stellar mass, implying a direct role for black holes in regulating the cold gas reservoir \citep[e.g.,][]{2024Natur.632.1009W}. The warm ionized phase is different: studies using \ion{C}{4} absorption find no clear correlation with SMBH mass, instead linking the warm gas to the specific star formation rate \citep[e.g.,][]{2024ApJ...970..115G}. The observational picture is thus mixed: robust correlations exist for hot and cold gas, yet the connection to the warm ionized medium remains weak or absent. Moreover, these studies are limited by sample size, selection effects, and the difficulty of measuring all gas phases uniformly \citep[e.g.,][]{2017ARA&A..55..389T}. This situation calls for a controlled theoretical investigation in which all gas phases are examined within a single framework.

A fundamental obstacle in interpreting these trends is that galaxy properties, including stellar mass, halo mass, environment, and SMBH mass, are tightly intercorrelated, making it difficult to isolate the unique role of the black hole. Hydrodynamical cosmological simulations provide an ideal tool for this task, enabling controlled experiments within a realistic cosmic context. Existing simulation studies of AGN feedback generally follow two strategies. The first traces causal mechanisms by analyzing how feedback energy heats or ejects gas \citep[e.g.,][]{2020MNRAS.499..768Z, 2021MNRAS.508.4667P, 2023MNRAS.518.5754R}. The second examines the endpoint statistical correlations that emerge from the long-term integrated effect of feedback \citep[e.g.,][]{2020MNRAS.493.1888T, 2020MNRAS.494..549T}. Within IllustrisTNG, recent work following this second approach has shown that the cold gas content of central galaxies is regulated by kinetic-mode AGN feedback, which ejects circumgalactic gas and suppresses cooling \citep[e.g.,][]{2020MNRAS.491.4462D, 2020MNRAS.499..768Z}. Comparative studies further demonstrate that while AGN feedback suppresses cold gas in massive galaxies across all models, the quantitative predictions depend sensitively on the specific subgrid implementation \citep[e.g.,][]{2020MNRAS.497..146D, 2021MNRAS.507.2383A}.

Despite this progress, a systematic census of the SMBH--multiphase gas connection is still missing. Existing simulation studies have typically focused on individual gas phases and have not rigorously controlled for the confounding effects of stellar and halo mass. It remains unclear how the statistical imprint of the SMBH varies across the full thermodynamic landscape of the CGM, from cold molecular and atomic gas through the warm ionized medium to the hot X-ray emitting halo, and how this imprint depends on galactocentric distance. A unified picture of how AGN feedback differentially couples to the various CGM phases is therefore lacking.

In this paper, we perform a systematic census of the endpoint statistical relationships between SMBH mass and multiphase gas content within IllustrisTNG. Using the TNG100 simulation at $z=0$, we measure the partial correlation between central SMBH mass and gas mass across four temperature-defined phases (cold, cool, warm, hot) and three spatial scales ($0.03R_{200}$, $0.15R_{200}$, $R_{200}$). We account for stellar mass and dark matter halo mass, thereby isolating the intrinsic association between $M_{\rm BH}$ and CGM gas. We further examine how this signal depends on galaxy type and local environment density. We find that, after removing these confounding factors, SMBH mass shows a significant negative partial correlation with cold gas on scales of $R_{200}$ and $0.15R_{200}$.

The paper is organized as follows. Section~\ref{sec:method} describes the simulation, sample selection, and partial correlation methodology. Section~\ref{sec:res} presents our results, progressing from raw correlations to intrinsic partial correlations, followed by the dependence on star formation activity and environment. Section~\ref{sec:discuss} compares our findings with observations and other simulations, discusses the physical interpretation in the context of the TNG feedback model, and outlines caveats and future directions. We summarize our conclusions in Section~\ref{sec:sum}.

\section{Methodology}
\label{sec:method}

\subsection{The IllustrisTNG100 Simulation}

We conduct our analysis using the IllustrisTNG100 cosmological magneto-hydrodynamical simulation \citep[TNG100;][]{2018MNRAS.475..624N, 2018MNRAS.475..648P, 2018MNRAS.475..676S, 2018MNRAS.477.1206N, 2018MNRAS.480.5113M}.\footnote{All IllustrisTNG data are publicly available at \url{https://www.tng-project.org}.}
We select TNG100 because it offers an optimal balance between sample size and resolution for our purposes: its volume provides a large number of well-resolved central galaxies for robust statistics, while its resolution reliably captures gas masses within both galactic and circumgalactic apertures.

The simulation evolves a periodic volume of approximately $(100\,c{\rm Mpc}/h)^3$ from redshift $z=127$ to $z=0$. The TNG100-1 run used here has a dark matter mass resolution of $7.5\times10^6\,M_\odot$ and a baryonic mass resolution of $1.4\times10^6\,M_\odot$. The gravitational softening length for dark matter and star particles is $0.74\,{\rm kpc}$ in physical units at $z=0$; for gas cells it is adaptive, with a minimum value comparable to this scale. Our analysis is based on the publicly available $z=0$ snapshot and the corresponding group and subhalo catalogs. We adopt the cosmological parameters of the TNG simulation, with $h=0.6774$.

A critical component of the TNG model for this study is its implementation of AGN feedback \citep{2017MNRAS.465.3291W}. This feedback operates in two distinct modes. In the ``thermal'' mode, active at high accretion rates, energy is deposited thermally to heat the surrounding gas. In the ``kinetic'' mode, which dominates at low accretion rates, momentum is injected via isotropic, stochastic winds. The kinetic mode is the primary mechanism in TNG for quenching massive galaxies, as it directly impacts and ejects circumgalactic gas.

\subsection{Sample Selection and Definitions}
\label{sec:sample}

\begin{deluxetable}{lccc|ccc|ccc|ccc}
\tabletypesize{\scriptsize}
\tablecaption{Number of galaxies with non-zero gas mass in each phase and aperture\label{tab:samples}}
\tablehead{
\colhead{Sample} & \multicolumn{3}{c}{Cold ($T<10^4$\,K)} & \multicolumn{3}{c}{Cool ($10^4\le T<10^5$\,K)} & \multicolumn{3}{c}{Warm ($10^5\le T<10^6$\,K)} & \multicolumn{3}{c}{Hot ($T\ge 10^6$\,K)} \\
\cline{2-4} \cline{5-7} \cline{8-10} \cline{11-13}
\colhead{} & \colhead{$0.03R_{200}$} & \colhead{$0.15R_{200}$} & \colhead{$R_{200}$} & \colhead{$0.03R_{200}$} & \colhead{$0.15R_{200}$} & \colhead{$R_{200}$} & \colhead{$0.03R_{200}$} & \colhead{$0.15R_{200}$} & \colhead{$R_{200}$} & \colhead{$0.03R_{200}$} & \colhead{$0.15R_{200}$} & \colhead{$R_{200}$}
}
\startdata
All galaxies     & 2523 & 4667 & 4858 & 4630 & 4981 & 5064 & 4475 & 5040 & 5072 & 4129 & 5064 & 5066 \\
Star-forming    & 2110 & 3219 & 3241 & 3343 & 3346 & 3346 & 3116 & 3346 & 3346 & 2609 & 3346 & 3346 \\
Green Valley    & 192  & 277  & 281  & 281  & 287  & 287  & 240  & 287  & 287  & 198  & 283  & 285  \\
Quenched        & 221 & 1171 & 1336 & 1006 & 1348 & 1431 & 1119 & 1407 & 1439 & 1322 & 1435 & 1435 \\
\enddata
\tablecomments{The full sample contains 5089 central galaxies with $M_* > 5\times10^9\,M_\odot$ and $M_{\rm BH} > 0$. Classifications follow Section~\ref{sec:sample}. For each phase and aperture, galaxies with zero gas mass after excluding star-forming cells are omitted from that entry.}
\end{deluxetable}

We focus on central galaxies at $z=0$ with stellar mass $M_* \ge 5\times10^9\,M_\odot$, where $M_*$ is the total mass of stellar particles bound to the subhalo within twice the stellar half-mass radius.  
We further require a non-zero total black hole mass ($M_{\rm BH} > 0$).
The final sample contains 5089 galaxies (3346 star-forming, 287 green valley, 1456 quenched). Key definitions are as follows.

\begin{enumerate}[label=(\roman*), align=left]

   \item Virial radius ($R_{200}$).
    $R_{200}$ is the radius of the sphere centered on the host dark matter halo within which the mean interior density equals 200 times the critical density of the Universe. 
We adopt three radial apertures throughout: $0.03R_{200}$, $0.15R_{200}$, and $R_{200}$, chosen to separate the galactic, inner halo, and full virial regions \citep[e.g.,][]{2019ApJ...884..169G}.
    
\item Gas mass measurement.
    For each galaxy, we sum the masses of all gas cells that are bound to the central subhalo and lie within each aperture. We exclude cells with non-zero star formation rate (SFR $> 0$) because IllustrisTNG assigns star-forming gas a non-thermodynamic temperature via an effective equation of state. This exclusion ensures that the remaining gas phases reflect genuine thermal properties of the CGM. 
Table~\ref{tab:samples} lists the effective sample sizes for each phase and aperture, after additionally omitting galaxies with zero gas mass in that bin.
    
\item Stellar mass ($M_*$) and dark matter halo mass ($M_{\rm dm}$).
    $M_*$ is measured within twice the stellar half-mass radius. $M_{\rm dm}$ is the total mass of dark matter particles bound to the subhalo. We use $M_{\rm dm}$ rather than the total halo mass $M_{200}$ as the halo mass control variable, because $M_{200}$ already contains the gas mass being measured, which would introduce a spurious self-correlation.        
    
    \item Black hole mass ($M_{\rm BH}$).
    $M_{\rm BH}$ is the sum of all black hole masses bound to the subhalo, dominated by the central black hole.
    
\item Star formation rate and galaxy classification.
    The SFR is the sum over all gas cells within twice the stellar half-mass radius. We classify galaxies by their offset from the star-forming main sequence at the corresponding redshift and stellar mass \citep{2021MNRAS.501.2210T, 2023MNRAS.519.1526P}:
    \[
    \begin{aligned}
    \text{Star-forming}&:\ \Delta\log\mathrm{SFR} > -0.5,\\
    \text{Green valley}&:\ -1.0 < \Delta\log\mathrm{SFR} \le -0.5,\\
    \text{Quenched}&:\ \Delta\log\mathrm{SFR} \le -1.0.
    \end{aligned}
    \]    
    
    \item Gas temperature and phases.
    Gas temperature in IllustrisTNG is an effective quantity derived from internal energy and electron abundance. After removing star-forming cells, we divide the remaining gas into four phases:
    \[
    \begin{aligned}
    \text{Cold}&:\ T < 10^4\,\mathrm{K},\\
    \text{Cool}&:\ 10^4 \le T < 10^5\,\mathrm{K},\\
    \text{Warm}&:\ 10^5 \le T < 10^6\,\mathrm{K},\\
    \text{Hot}&:\ T \ge 10^6\,\mathrm{K}.
    \end{aligned}
    \]

    \item Local environment density.
    The local density $\Sigma_5$ is defined as $\Sigma_5 = 5/(\pi r_5^2)$, where $r_5$ is the projected distance to the fifth nearest neighbor among the full sample of central galaxies, accounting for periodic boundary conditions. 
Only central galaxies are included in the neighbor pool, so that $\Sigma_5$ traces the large-scale environment rather than the host halo mass.
The sample is split into three equal groups (low, medium, high) by $\Sigma_5$ terciles.
    
\end{enumerate}

\subsection{Partial Correlation Analysis}

To isolate the intrinsic connection between gas mass and black hole mass from the confounding effects of stellar and halo mass, we perform the following procedure separately for each galaxy type (star-forming, green valley, quenched). 
The log-linear form is standard for galaxy scaling relations and is justified by their power-law behavior over the mass range probed. For a given type, we fit two ordinary least-squares linear models:
\begin{eqnarray}
\log M_{\rm gas} &=& a_1 \log M_* + b_1 \log M_{\rm dm} + c_1 , \nonumber \\
\log M_{\rm BH}   &=& a_2 \log M_* + b_2 \log M_{\rm dm} + c_2 .
\end{eqnarray}
Residuals for each galaxy are computed from the regression coefficients of its own type. The Spearman rank correlation between the two sets of residuals within a given type gives the type-specific partial correlation; the full-sample partial correlation is then obtained by pooling the type-specific residuals across all three types before computing the correlation. 
This partial correlation measures the association between gas and black hole mass after removing linear dependencies on $M_*$ and $M_{\rm dm}$.

We estimate 95\% confidence intervals for $\rho$ using bootstrap resampling with 1000 iterations. Statistical significance is marked as \ostar \ for $p<0.05$, \dstars \ for $p<0.01$, and \tstars \ for $p<0.001$; unmarked values have $p \ge 0.05$. The same group-specific regression and pooling procedure is applied to subsamples split by local environment density (low, medium, high).

\section{Results}
\label{sec:res}

\subsection{Raw Gas-BH Correlations}

\begin{figure*}[htp]
\centering
\includegraphics[width=0.98\textwidth]{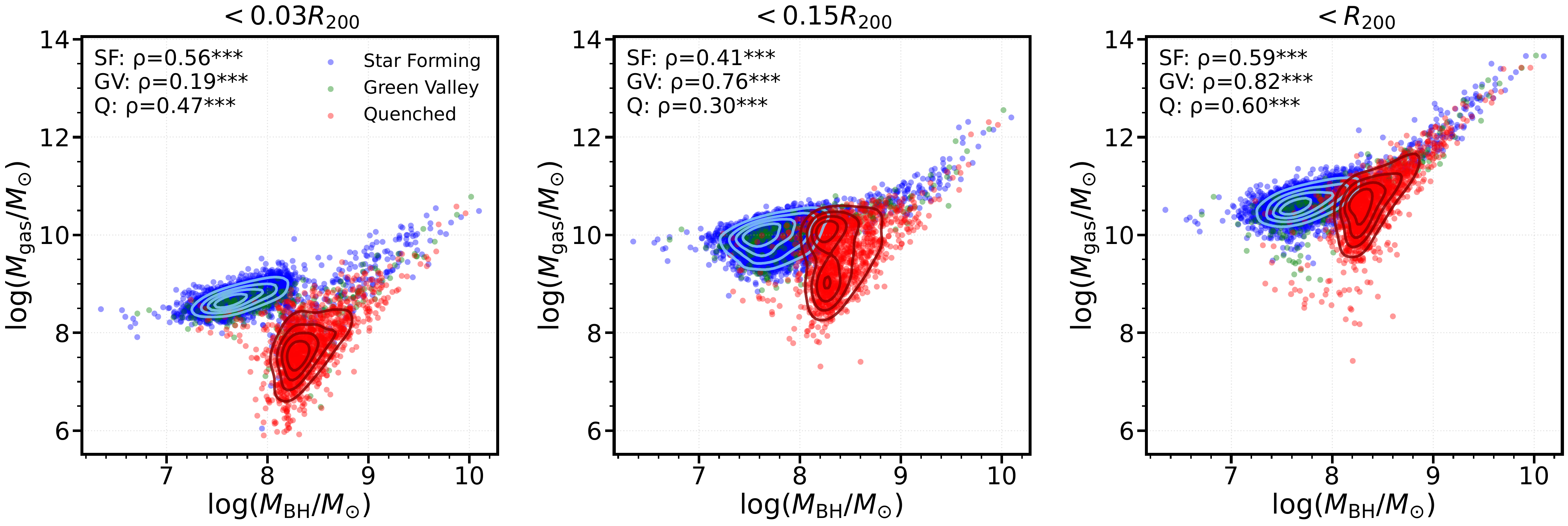}
\caption{Raw gas--BH correlations at three radial scales. Total gas mass (excluding star-forming gas cells) within $0.03 R_{200}$ (left), $0.15 R_{200}$ (middle), and $R_{200}$ (right) is plotted against black hole mass. Colors indicate galaxy type: star-forming (blue), green valley (green), and quenched (red). Contours (light blue for star-forming, dark red for quenched) show the 2D density distribution; green valley galaxies are not shown with contours due to their small sample size. Spearman correlation coefficients $\rho$ for star-forming (SF), green valley (GV), and quenched (Q) galaxies are given in each panel; \tstars \ indicates $p<0.001$.
\label{fig-1}}
\end{figure*}

Figure~\ref{fig-1} shows the raw correlations between total gas mass (excluding star-forming gas cells) and black hole mass at three radial scales. Star-forming and quenched galaxies form two distinct populations in the gas--BH plane, indicating that they follow different relations. The separation is strongest at $0.03R_{200}$ and diminishes toward $R_{200}$, indicating that the gas content of different galaxy types converges at larger radii. This motivates a separate partial correlation analysis for each galaxy type.

Green valley galaxies show a weak but significant positive correlation at the innermost scale ($\rho = 0.19$, $p<0.001$), and at $0.15R_{200}$ and $R_{200}$ their raw correlations reach $\rho = 0.76$ and $0.82$, though the small sample size ($N=287$) warrants caution. All raw correlations are positive, with star-forming and quenched galaxies spanning moderate to strong coefficients ($\rho = 0.3$--$0.6$). These raw trends are largely driven by the common dependence on stellar and halo mass, motivating the partial correlation analysis that follows.

\subsection{Intrinsic Correlations: Partial Analysis}

\begin{figure*}[htbp]
\centering
\includegraphics[width=0.98\textwidth]{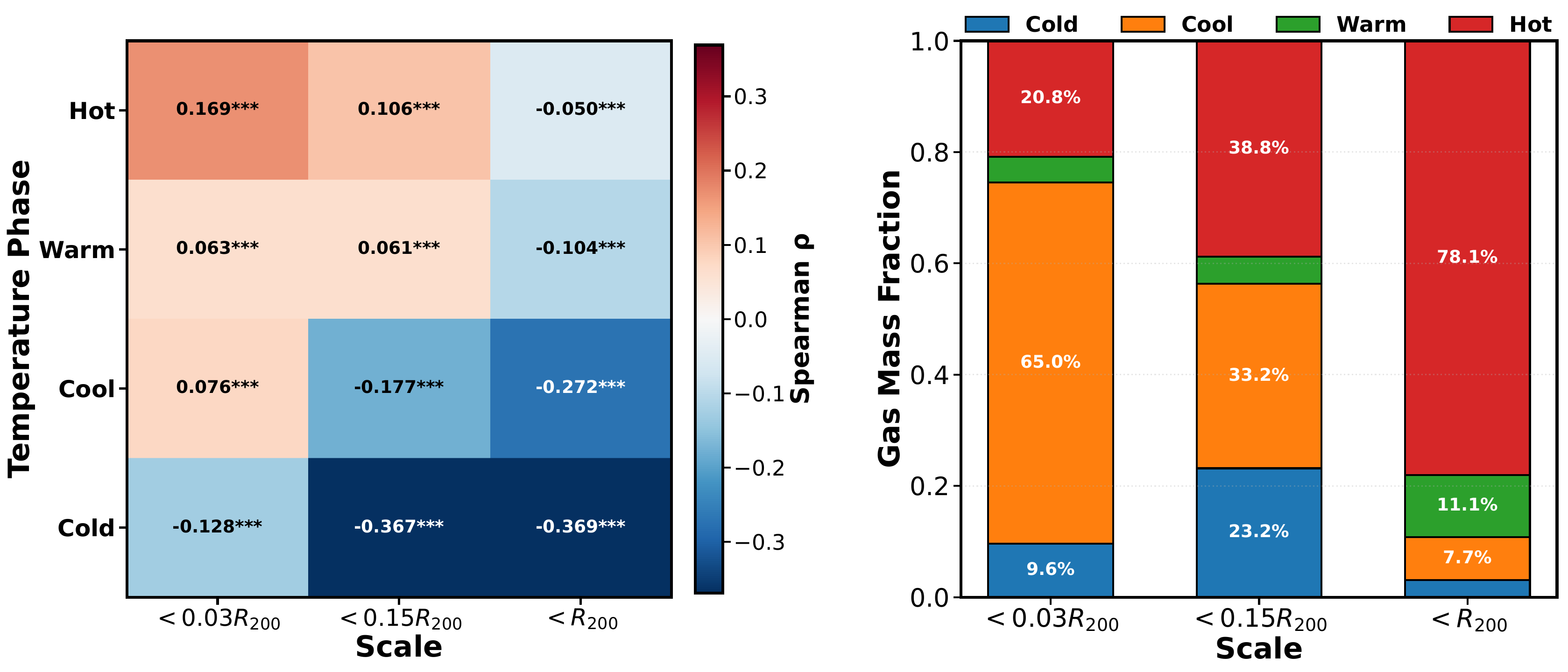}
\caption{Intrinsic partial correlations and gas mass fractions. Left panel: partial correlation heatmap for the full sample, after accounting for $\log M_*$ and $\log M_{\rm dm}$ via group-specific regressions. Colors represent Spearman's $\rho$; numbers in cells are $\rho$ values with significance levels (\tstars \ for $p<0.001$). Right panel: stacked bar chart of gas mass fractions in each temperature phase (Cold: $T<10^4$\,K, Cool: $10^4\le T<10^5$\,K, Warm: $10^5\le T<10^6$\,K, Hot: $T \ge 10^6$\,K) within the three radial bins, after excluding star-forming gas cells. Percentages are shown for phases with $>5\%$ contribution.
\label{fig-2}}
\end{figure*}

The left panel of Figure~\ref{fig-2} shows the intrinsic gas--BH connection after removing the linear effects of $M_*$ and $M_{\rm dm}$ via group-specific regressions. The strongest signal is a moderate negative partial correlation for cold gas on scales of $0.15R_{200}$ and $R_{200}$ ($\rho \approx -0.37$). Cool gas shows a weaker but significant anti-correlation on the scale of $R_{200}$, while on the $0.03R_{200}$ scale the signals are weak or absent for all phases. Hot gas shows only a weak positive correlation on small scales that vanishes on larger scales; warm gas shows no substantial correlation. The contrast between the raw and partial correlations confirms that the apparent gas--BH connection largely arises from the common dependence on halo mass, whereas the intrinsic signature of AGN feedback appears as a deficit of cold gas on large scales, with a weaker extension to the cool phase.

The right panel of Figure~\ref{fig-2} shows the stacked gas mass fractions. Within $R_{200}$, the hot gas dominates, accounting for roughly $78\%$ of the total, while cool and cold gas together make up only about $11\%$. Despite this low abundance, the anti-correlations for cold gas are statistically robust, because the partial correlation operates on residuals rather than on absolute gas masses and the sample is large. 
We caution that the stacked fractions mask the limited resolution of cold gas in the inner zone; this is discussed further in Section~\ref{sec:discuss_caveats}.
The dominant hot gas, by contrast, shows no intrinsic anti-correlation on large scales, reinforcing the interpretation that the raw hot gas--BH signal is a halo-mass-driven artifact.

\begin{figure}[htbp]
\centering
\includegraphics[width=0.47\textwidth]{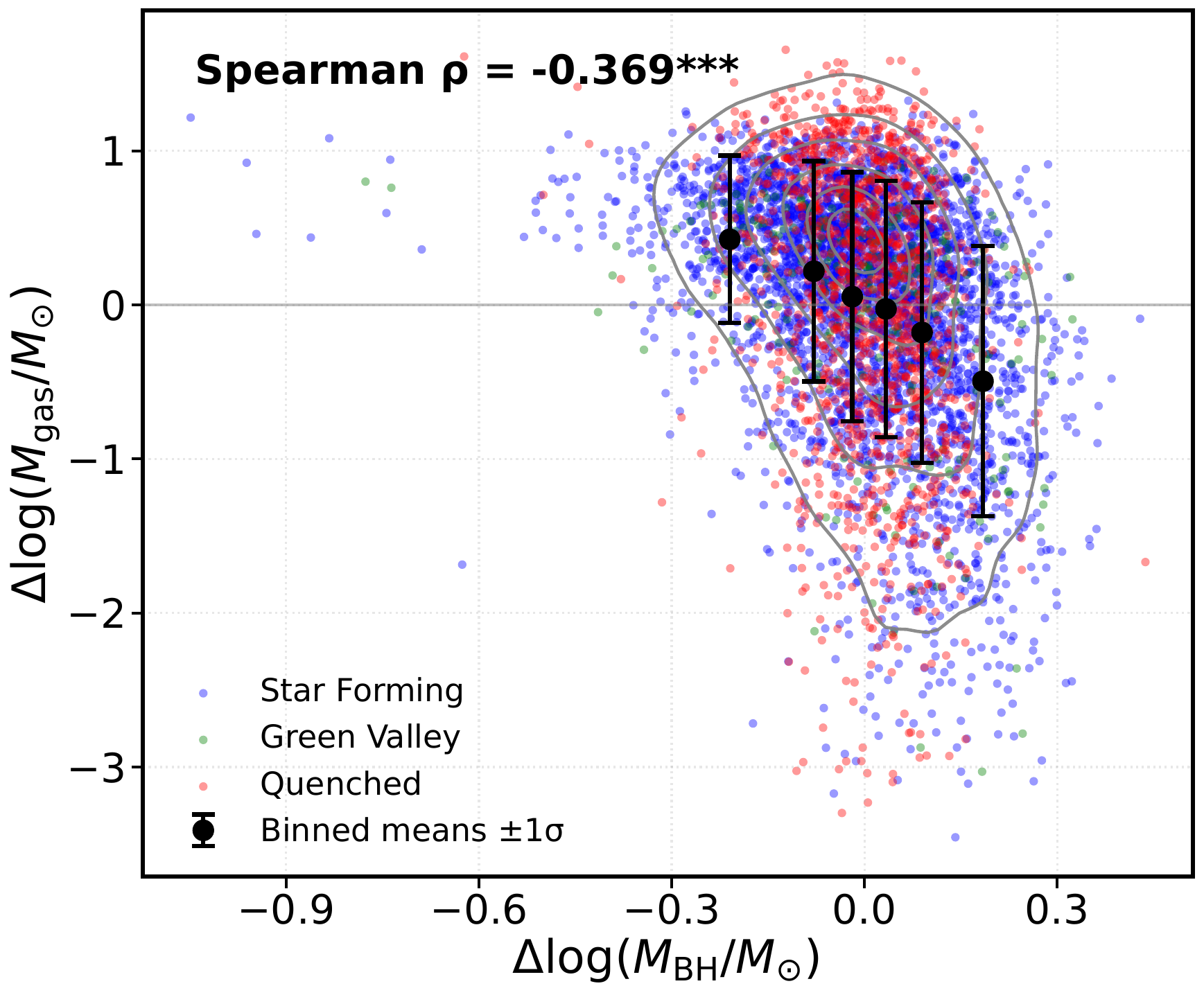}
\caption{Residual plot for the strongest intrinsic anti-correlation from Figure~\ref{fig-2} (cold gas within $R_{200}$), showing gas mass residuals against BH mass residuals after removing $\log M_*$ and $\log M_{\rm dm}$. Colors follow Figure~\ref{fig-1}. Gray contours show the 2D density distribution. Black circles with error bars are binned means $\pm1\sigma$ (six equal-frequency bins). The horizontal line marks zero gas residual. The Spearman partial correlation coefficient is given in the upper left.
\label{fig-3}}
\end{figure}

We now inspect the strongest of these signals, the anti-correlation for cold gas within $R_{200}$ ($\rho = -0.369$), in more detail. Figure~\ref{fig-3} shows the corresponding residual plot. The residual plane reveals an asymmetric pattern: galaxies with negative BH residuals (black holes less massive than expected) show gas residuals that scatter around slightly positive values, whereas those with positive BH residuals (over-massive black holes) show systematically depleted gas. This threshold-like behavior is also visible in the binned means and density contours. Star-forming, green valley, and quenched galaxies are all present across the full range of BH residuals, with similar fractions in the negative and positive regimes. 
Both quenched and star-forming galaxies exhibit extreme cold-gas depletion in the positive residual region, indicating that over-massive black holes are associated with cold-gas removal across the full galaxy population, consistent with AGN feedback operating progressively rather than as a binary switch.

The absence of a single linear trend across the full BH residual range, combined with the similar galaxy-type fractions in the positive and negative residual regimes, indicates that the gas--BH connection is not a simple monotonic relation. 
Rather than a sharp break at a fixed mass threshold, the pattern suggests a progressive effect: once the black hole becomes over-massive relative to its host, cold-gas depletion increases in a continuous manner, consistent with the cumulative impact of AGN feedback over time.

\subsection{Star Formation Activity Dependence}
\label{sec:res_sf}

\begin{figure*}[htp]
\centering
\includegraphics[width=0.98\textwidth]{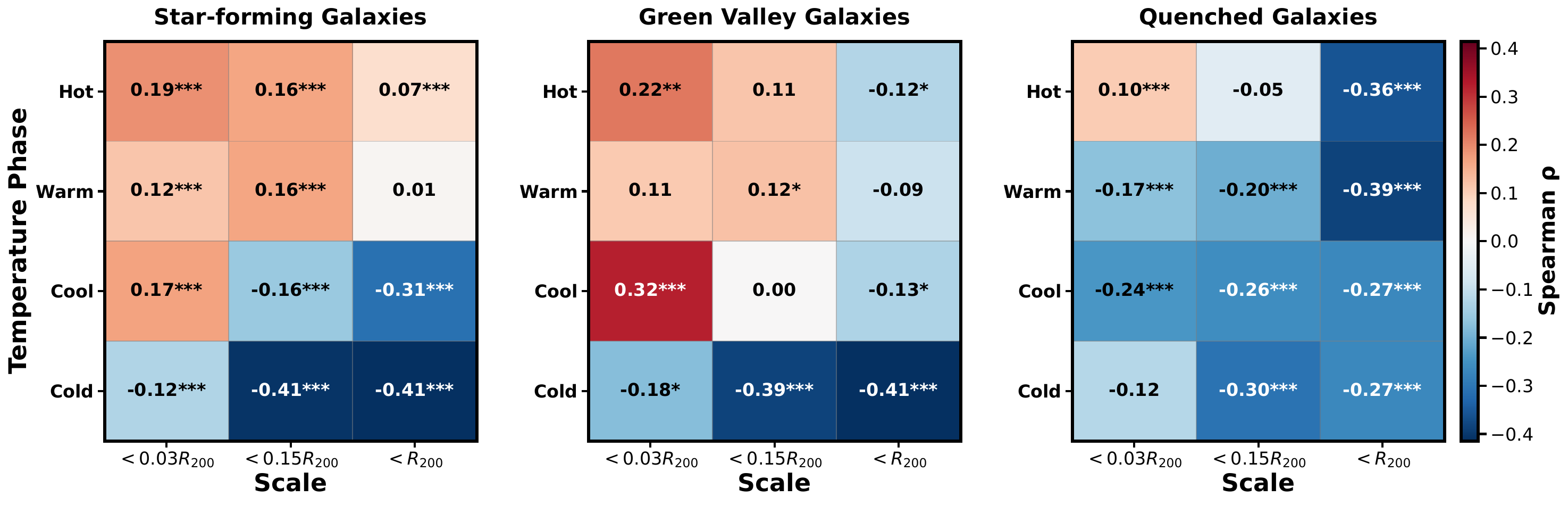}
\caption{Dependence on star formation activity. Left, middle, and right panels show partial correlation heatmaps for star-forming, green valley, and quenched galaxies, respectively. Significance levels follow the convention of Fig.~\ref{fig-2}.
\label{fig-4}}
\end{figure*}

Figure~\ref{fig-4} splits the partial correlation analysis by galaxy type. The intrinsic gas--BH anti-correlation depends clearly on star formation activity.

In star-forming galaxies, hot and warm gas show weak positive partial correlations on small and intermediate scales ($0.03R_{200}$ and $0.15R_{200}$), suggesting that the inner CGM in these systems is still governed by host halo properties rather than by the central black hole. At $R_{200}$, these correlations vanish or become very weak, indicating that even the hot halo is not systematically linked to $M_{\rm BH}$. The negative imprint of AGN feedback appears only in the cold and cool phases: a weak anti-correlation within $0.03R_{200}$ strengthens to $\rho = -0.41$ within $0.15R_{200}$ and $R_{200}$. This outward strengthening indicates that the imprint of AGN feedback on cold gas is most pronounced at circumgalactic scales.

Green valley galaxies show a mixed pattern. Within $0.03R_{200}$, cool gas exhibits a significant positive correlation ($\rho = 0.32$, $p<0.001$) while cold gas shows a weak negative correlation ($\rho = -0.18$, $p<0.05$). The positive cool gas signal may trace residual inflow that still feeds the central black hole. 
Moving outward, the positive cool-gas correlation vanishes, while cold gas develops a significant anti-correlation with $\rho = -0.39$ within $0.15R_{200}$ and $-0.41$ within $R_{200}$, indicating that feedback already dominates the gas content in the CGM of these transitioning galaxies.

Quenched galaxies show weaker cold gas anti-correlations than star-forming or green valley systems, while their cool gas anti-correlations remain significant and are comparable across all three scales. At $0.03R_{200}$, the cold gas correlation is not significant ($\rho = -0.12$), consistent with the weak signals seen in the full sample on this scale. In contrast to star-forming and green valley galaxies, where warm and hot gas show only weak positive or null correlations, quenched galaxies exhibit significant negative correlations in the warm phase at all three scales and in the hot phase within $R_{200}$. This shift toward warm- and hot-phase anti-correlations may reflect cumulative heating of the CGM by past AGN feedback.

In summary, the intrinsic gas--BH anti-correlation varies systematically with galaxy type: it is scale-dependent in star-forming and green valley galaxies, and weakens in the cold phase while shifting toward warm and hot phases in quenched systems. This progression points to a radially increasing feedback imprint, consistent with the outward-directed impact of AGN feedback.

\subsection{Local Density Effect}

\begin{figure*}[htbp]
\centering
\includegraphics[width=0.98\textwidth]{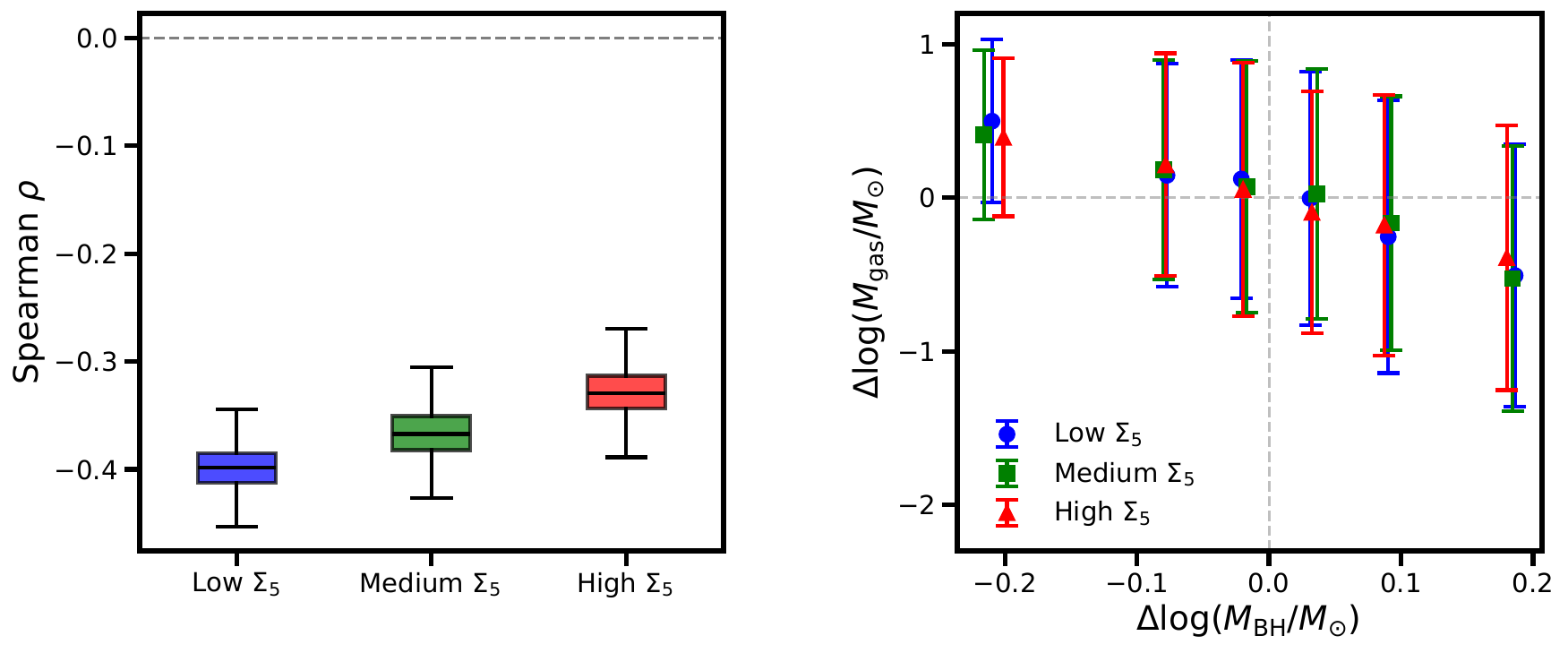}
\caption{Local density effect. Left panel shows the bootstrap distributions of the partial correlation coefficient for cold gas within $R_{200}$ for low, medium, and high local density groups (based on $\Sigma_5$), computed using the same group-specific regression and pooling procedure as in Fig.~\ref{fig-2}. Boxes mark the interquartile range, horizontal lines inside the boxes indicate the median, and whiskers extend to the 5th and 95th percentiles. The dashed line marks zero correlation. Right panel shows binned residual means for the same density groups. Error bars denote $\pm1\sigma$ in each bin.
\label{fig-5}}
\end{figure*}

We test whether the intrinsic gas--BH anti-correlation depends on environment, using the projected surface density to the fifth nearest neighbor, $\Sigma_5$. We focus on cold gas within $R_{200}$, the strongest signal from Figure~\ref{fig-2}. For each density bin (low, medium, high), we compute the partial correlation and its bootstrap confidence interval following the same procedure as in Section~\ref{sec:method}. The left panel of Figure~\ref{fig-5} shows the resulting distributions.
The anti-correlation is strongest in low-density environments ($\rho \approx -0.40$) and weakest in high-density environments ($\rho \approx -0.33$). While the full bootstrap distributions overlap, the interquartile ranges are clearly separated between the low- and high-density groups, indicating a modest but discernible trend with local density. The binned residuals (right panel) are nearly identical across groups.

The anti-correlation thus persists across all density bins, but a weak trend is present: the signal weakens from low to high local density. This suggests that environmental processes at higher densities may partially dilute the intrinsic AGN feedback signature. In contrast to the clear dependence on star formation activity (Section~\ref{sec:res_sf}), the environmental trend is modest, indicating that the cold-gas--BH connection is largely preserved across different environments.

\section{Discussion}
\label{sec:discuss}

\subsection{Comparison with Observational Constraints}
\label{sec:discuss_obs}

Our TNG100 analysis provides testable predictions for current and future observational campaigns. Here we compare our findings with observations across the four gas phases.

\textbf{Hot gas ($T \ge 10^6$\,K).}
X-ray observations of groups and clusters show that $M_{\rm BH}$ correlates tightly with both the total gravitating mass of the host halo \citep[e.g.,][]{2018ApJ...852..131B, 2019MNRAS.488L.134L, 2019ApJ...875..141P} and the properties of the hot atmosphere, such as its X-ray temperature and luminosity \citep[e.g.,][]{2019ApJ...884..169G}. Our analysis extends these findings by isolating the intrinsic connection: after removing the linear effects of $M_*$ and $M_{\rm dm}$, the partial correlation for hot gas is consistent with zero at $R_{200}$ in the full sample (Figure~\ref{fig-2}), indicating that the hot gas reservoir is set primarily by the halo potential. However, a significant anti-correlation emerges in quenched galaxies (Figure~\ref{fig-4}), suggesting that cumulative AGN heating of the CGM becomes detectable only after star formation has ceased.

\textbf{Cold gas ($T < 10^4$\,K).}
\citet{2024Natur.632.1009W} recently analysed the ALFALFA \ion{H}{1} survey and used a similar partial correlation approach to show that the \ion{H}{1}-to-stellar mass fraction is more strongly correlated with $M_{\rm BH}$ than with $M_*$. 
Our results are consistent with theirs: cold gas shows a moderate negative partial correlation with $M_{\rm BH}$ at $0.15R_{200}$ and $R_{200}$ ($\rho \approx -0.37$; Figure~\ref{fig-2}), after controlling for both $M_*$ and $M_{\rm dm}$. The agreement supports the picture that AGN feedback depletes the cold circumgalactic reservoir.

\textbf{Cool and warm gas ($10^4 \le T < 10^6$\,K).}
Observationally, the warm phase ($10^5$--$10^6$\,K), traced by ions such as \ion{C}{4} and \ion{O}{6}, shows no significant partial correlation with $M_{\rm BH}$ \citep[e.g.,][]{2024ApJ...970..115G}. TNG100 gives the same null result for warm gas. The cool phase ($10^4$--$10^5$\,K) shows a significant negative partial correlation ($\rho \approx -0.27$ within $R_{200}$; Figure~\ref{fig-2}). This temperature range, corresponding to the peak of the cooling curve, is expected to be observable in absorption via low-ionization metal lines (e.g., \ion{Si}{4}, \ion{C}{3}, \ion{Mg}{2}) and \ion{H}{1} Ly$\alpha$. Large statistical samples of cool CGM gas are still lacking. Upcoming absorption-line surveys with facilities such as DESI and WEAVE will provide the data needed to search for this anti-correlation, provided that stacking analyses and careful ionization modeling are used to isolate the cool phase from warmer gas.

\subsection{Model Dependence and Comparison with Other Simulations}
\label{sec:discuss_sim}

Our results are specific to the AGN feedback prescription in IllustrisTNG. 
In TNG, the kinetic feedback mode becomes dominant above a characteristic black hole mass of $\sim10^8\, M_\odot$ \citep{2017MNRAS.465.3291W}. The threshold-like depletion pattern in Fig.~\ref{fig-3} is related to this scale but is not confined to the most massive galaxies; the partial correlation captures a continuous trend consistent with cumulative AGN feedback. This picture, in which $M_{\rm BH}$ traces the cumulative impact of feedback, is consistent with previous analyses identifying $M_{\rm BH}$ as the most predictive parameter for gas depletion and quenching \citep{2022MNRAS.512.1052P, 2023ApJ...944..108B}.

No uniform partial correlation analysis across different simulations exists yet, but we can assess the robustness of our approach by comparing with existing studies within TNG. For the hot gas phase, earlier TNG work \citep{2020MNRAS.494..549T} showed that the X-ray properties of hot atmospheres are tightly correlated with galaxy mass, with the imprint of SMBH feedback becoming visible only at the transition scale between star-forming and quenched galaxies. Our partial correlation analysis extends this picture: by controlling for both $M_*$ and $M_{\rm dm}$, we find that the hot gas partial correlation is consistent with zero at $R_{200}$ (Figure~\ref{fig-2}), indicating that the hot gas reservoir is set primarily by the host halo rather than by the central black hole. For the cold gas phase, our finding of a significant negative partial correlation ($\rho \approx -0.37$) is consistent with TNG-based analyses showing that kinetic-mode feedback suppresses cold gas and is closely tied to $M_{\rm BH}$ \citep{2020MNRAS.491.4462D, 2020MNRAS.499..768Z}.

These consistencies within TNG motivate a broader comparison. EAGLE uses a purely thermal feedback mode that heats gas and suppresses cooling \citep{2015MNRAS.446..521S}. This mechanism primarily counteracts gas accretion in the inner halo, so any cold-gas deficit is expected to be concentrated at small radii. The outward strengthening of the anti-correlation seen in TNG (Figure~\ref{fig-2}) would therefore not be expected in a model like EAGLE, where feedback is purely thermal. In contrast, SIMBA includes a jet mode that transports energy to large radii and heats the outer halo \citep{2019MNRAS.486.2827D}. Comparative studies indicate that SIMBA produces a hotter, more volume-filling warm-hot phase than TNG \citep{2021MNRAS.507.2383A}, which may lead to stronger intrinsic anti-correlations in the warm and hot phases than those we find in TNG.

These comparisons highlight the model dependence of the multiphase gas--BH connection. Applying the partial correlation framework developed here uniformly to other cosmological simulations would directly test how different feedback implementations shape the CGM. Extending the analysis to TNG50 and TNG300 would further test the robustness of our results against numerical resolution and cosmic variance. The TNG predictions presented here offer a baseline for such work, and the agreement with existing TNG analyses supports the robustness of our approach.

\subsection{Caveats and Future Prospects}
\label{sec:discuss_caveats}

Several limitations of our analysis should be kept in mind.

First, our results are specific to a single simulation (IllustrisTNG100) and use fixed temperature thresholds to define gas phases. Observational tracers are sensitive to specific ionization states and abundances, so forward modeling with photoionization and radiative transfer will be required to connect our predictions directly to observable quantities.

Second, the cold gas within $0.03R_{200}$ is poorly resolved for a substantial fraction of the sample, and the stacked fractions in Fig.~\ref{fig-2} mask this variation. The weak or null partial correlations on this scale should therefore be interpreted with caution; our main conclusions rest on the better-resolved signals at larger radii.

Third, the $z=0$ snapshot provides only a static endpoint, and our partial correlation analysis captures the cumulative effect of AGN feedback without distinguishing between thermal and kinetic modes. Tracking individual galaxies across cosmic time would establish a more direct causal link between SMBH growth and gas removal, and help separate the contributions of different feedback channels. Extending the partial correlation framework to TNG50 and TNG300 would test how numerical resolution and cosmic variance affect our results, particularly within $0.03R_{200}$ where the current resolution is limited.

Finally, the cold gas phase in our analysis includes both atomic and molecular gas, whereas the most directly comparable observational tracer is the 21\,cm line of \ion{H}{1}. Future observations capable of resolving the cold circumgalactic component cleanly from the star-forming interstellar medium will be essential for a definitive comparison.

Despite these limitations, this work provides a systematic, multi-phase baseline for the SMBH--CGM connection in IllustrisTNG and offers clear predictions for future multiwavelength surveys.

\section{Summary and Conclusions}
\label{sec:sum}

We have performed a systematic partial correlation analysis of the multiphase gas--black hole connection in the IllustrisTNG100 simulation at $z=0$. After accounting for $M_*$ and $M_{\rm dm}$, we isolated the intrinsic imprint of SMBH mass on cold, cool, warm, and hot gas across three radial scales, and examined how this imprint varies with galaxy type and environment. Our main conclusions are as follows:

\begin{enumerate}

\item The raw positive correlations between total gas mass and $M_{\rm BH}$ are largely driven by the shared dependence on halo mass. After removing this dependence via group-specific regressions, a significant negative partial correlation emerges for cold gas on scales of $0.15R_{200}$ and $R_{200}$ ($\rho \approx -0.37$), with a weaker signal for cool gas, and no significant signal on the $0.03R_{200}$ scale. Warm gas shows no substantial intrinsic correlation, while hot gas exhibits only a weak positive correlation on the smallest scale ($0.03R_{200}$) that vanishes at $R_{200}$.

\item The strongest anti-correlation is found for cold gas within $R_{200}$ ($\rho = -0.369$), with the signal within $0.15R_{200}$ being nearly identical ($\rho = -0.367$). The residual plane reveals a threshold pattern: galaxies with over-massive black holes display systematic cold-gas depletion, consistent with a picture in which AGN feedback progressively depletes cold gas once the SMBH becomes over-massive relative to its host halo.

\item The anti-correlation varies with galaxy type. Star-forming galaxies display a pronounced outward strengthening of the signal in cold gas. Green valley galaxies show a mixed pattern, with a positive cool-gas correlation at the centre that vanishes at intermediate radii and turns negative at larger scales, consistent with a transitional stage where fueling and feedback coexist. Quenched galaxies show weaker cold-gas anti-correlations, but warm and hot gas become significantly anti-correlated within $R_{200}$, suggesting cumulative heating of the CGM by past AGN feedback.

\item The anti-correlation for cold gas persists across all environments but, unlike its clear dependence on star formation activity, shows only a weak trend with local density ($\rho \approx -0.40$ to $-0.33$). The signal is strongest in low-density regions, where the AGN feedback signature is least diluted by environmental processes.

\end{enumerate}

These results establish a unified, multiphase view of the SMBH--CGM connection in IllustrisTNG. The agreement with independent TNG analyses supports the robustness of the partial correlation approach. The distinct radial and thermodynamic signatures identified here, in particular the outward strengthening of the anti-correlation for cold gas and the emergence of warm- and hot-phase anti-correlations in quenched systems, are consistent with AGN feedback and can be tested with future multiwavelength surveys. Extending this partial correlation framework to other simulations and higher redshifts will be essential for assessing the generality of these findings and for understanding how AGN feedback shapes the circumgalactic ecosystem.

\begin{acknowledgments}
We thank the anonymous referee for helpful comments that improved the quality of this work.
This work is supported by the National SKA Program of China No. 2025SKA0150103, National Natural Science Foundation of China under Nos. 12550002, 12133008, 12221003, 11890692, 12233005. We acknowledge the science research grants from the China Manned Space Project with No. CMS-CSST-2021-A04 and No. CMS-CSST-2025-A10, the Fujian Provincial Natural Science Foundation of China (Grant No. 2024J08001), and the Natural Science Foundation of Xiamen, China (No. 3502Z202472007). 
\end{acknowledgments}

\software{numpy \citep{2020Natur.585..357H},
          scipy \citep{2020NatMe..17..261V},
          matplotlib \citep{2007CSE.....9...90H},
          pandas \citep{2022zndo...3509134T},
          astropy \citep{2013A&A...558A..33A,2018AJ....156..123A,2022ApJ...935..167A}
          }

\bibliography{ms}{}
\bibliographystyle{aasjournal}

\end{CJK*}
\end{document}